\input{aipcheck}

\documentclass[
    ,final            
  ]
  {aipproc}

\usepackage{epsfig}
\usepackage{amssymb,amsmath}

\layoutstyle{8x11single}

\begin{document}

\title{Multi-Higgs doublet models with local $U(1)_H$ gauge \\ symmetry 
and neutrino physics therein}

\classification{12.60.Fr, 14.65.Ha, 14.60.St}
\keywords      {Higgs, Two Higgs doublet model, Higgs flavor, Top quark, Neutrinos}

\author{P. Ko}{
  address={School of Physics, KIAS, Seoul 130-722, Korea}
}

\author{Yuji Omura}{
 address={Physik Department T30, Technische Universit\"{a}t M\"{u}nchen,
James-Franck-Stra$\beta$e, 85748 Garching, Germany}
}

\author{Chaehyun Yu}{
  address={School of Physics, KIAS, Seoul 130-722, Korea}
}

\begin{abstract}
Multi-Higgs doublet models appear in many interesting extensions of the standard 
model (SM). But they suffer from Higgs-mediated flavor changing neutral current (FCNC)  
problem which is very generic.  In this talk, I describe that this problem can be resolved 
or mitigated if we introduce local $U(1)_H$  Higgs flavor gauge symmetry. As examples, 
I describe chiral $U(1)_{H}$ models where the right-handed up-type quarks also carry 
$U(1)_H$ charges and discuss the top forward-backward asymmetry (FBA) and 
$B\rightarrow D^{(*)} \tau \nu$ puzzle.  Next I describe the two-Higgs doublet models 
where the usual $Z_2$ symmetry is implemented to $U(1)_H$ and show how the Type-I 
and  Type-II models are extended. One possible extension of Type-II has the same 
fermion contents with the leptophobic $E_6$ $Z^{'}$ model by Rosner,  and I discuss 
the neutrino sector in this model briefly.
\end{abstract}

\maketitle


\section{Introduction}
The long-sought-for Higgs boson has been finally discovered at the LHC. 
Still the Higgs sector is the least understood part of the standard model (SM), 
both theoretically and experimentally. 
There are still many questions about the electroweak symmetry breaking (EWSB) sector: 
\begin{itemize}
\item Is it exactly the same as the SM Higgs boson or is there any (slight) deviation from 
the SM Higgs boson ? 
\item How many Higgs doublets are  there ?  
\item Is there any singlet scalar that could mix with the SM Higgs boson ?
\item Are there Higgs multiplets with weak isospin larger than $1/2$ ?
\end{itemize}
The upcoming LHC @ 13, 14 TeV and the future linear collider will provide us with 
invaluable information about the origin of electroweak symmetry breaking (EWSB) 
and (partial) answers to the questions listed above.  

Among many extensions of the SM Higgs sector, multu-Higgs doublet models are 
well motivated in various extensions beyond the SM (BSM).
In this talk, I first discuss how my collaborators and I came up with the idea of Higgs 
flavor in the context  of chiral $U(1)^{'}$ flavor models with multi-Higgs doublets 
invented for the top forward-backward asymmetry at the Tevatron.  
Then I describe our proposal to implement the softly broken $Z_2$ symmetry in 2 Higgs 
doublet models (2HDMs) to a spontaneously broken $U(1)_H$ symmetry, 
and show how the usual Type-I and Type-II 2HDMs are generalized into  new 2HDMs 
with $U(1)_H$ symmetry.  
In particular one extension of the Type-II 2HDM  is exactly the same as the leptophobic 
$Z^{'}$ models derived from $E_6$ by J.L. Rosner~\cite{Rosner:1996eb}.
In this model, there are new sterile neutrinos whose mass matrix is fixed by gauge 
quantum  numbers of the SM fermions and the new sterile neutrinos, which I 
briefly touch upon. 

\section{Chiral $U(1)^{'}$ flavor models for the  top FBA}

\subsection{Motivations}

The top forward-backward asymmetry ($A_\textrm{FB}^t$) has been one of the most 
interesting observables recently, since there have been some discrepancies between 
theoretical predictions in the standard model (SM) and experimental results at the 
Tevatron.   The most recent measurement for $A_\textrm{FB}^t$ at CDF is 
$A_\textrm{FB}^t=0.162\pm 0.047$ in the letpon+jets channel with a full
set of data~\cite{cdfnew}, which is consistent with
the previous measurements at CDF and D0 within uncertainties~\cite{oldafb}. 
The SM predictions are between $0.06$ and $0.09$~\cite{smafbac,smafb},
so that the deviation is around $2 \sigma$.

If the discrepancy in $A_\textrm{FB}^t$ is generated by new physics beyond the SM,
the new physics might be tested at the LHC. 
One of the good measurements is the charge asymmetry $A_C^y$, which is 
defined by the difference of numbers of events with the positive and negative 
$\Delta |y|=|y_t|-|y_{\bar{t}}|$ divided by their sum.
The current values for $A_C^y$ are
$A_C^y=-0.018\pm 0.028\pm 0.023$ at ATLAS~\cite{atlasacy} and 
$A_C^y=0.004\pm 0.010\pm 0.012$ at CMS~\cite{cmsacy}, respectively, 
which are consistent with the SM prediction $\sim 0.01$~\cite{smafbac}. 
Another interesting observable at the LHC is the cross section for
the same-sign top-quark pair 
production, $\sigma^{tt}$, which is not allowed in the SM. 
The current upper bound on $\sigma^{tt}$ is about 0.39 pb at 95 \% 
C.L..~\cite{Chatrchyan:2012sa} and 
2 pb or 4 pb at ATLAS depending on the model~\cite{atlassame}.   
Some models which were proposed to account for $A_\textrm{FB}^t$ at the Tevatron,
predict large $A_C^y$ and/or $\sigma^{tt}$ so that they are already disfavored
by present experiments at the LHC.  However the story is not that simple if there is 
a new chiral gauge boson, which is a main theme of this section.

\subsection{The original $Z^{'}$ model by Jung {\it et al.} \cite{zprime}}
Let us consider a $Z'$ model first proposed by Jung, Murayama, Pierce and 
Wells ~\cite{zprime}, who assume that there is a flavor changing $Z'$ couplings 
to the right-handed (RH) $u$ and $t$ quarks:
\begin{equation}
{\cal L} = - g_X Z_\mu^{'} \left[  \overline{t_R} \gamma^\mu u_R + H.c. \right] .
\end{equation}
The $t$-channel exchange of $Z'$ leads to the Rutherford peak in the forward direction
and generates the desired amount of the top FBA if $Z'$ is around $150-250$ GeV and
$g_X$ is not too small. Here $Z'$ is assumed to couple only to the right-handed (RH) 
quarks in order to evade the strong bounds from the FCNC processes such as 
$K^0 - \overline{K^0}$,
$B^0_{d(s)} - \overline{B_{d(s)^0}}$ mixings and $B \rightarrow X_s \gamma$, etc.. 
Such a  light $Z^{'}$ should be leptophobic in order to avoid the strong bounds 
from the Drell-Yan  processes. Therefore the original $Z'$ model by Jung et al.
\cite{zprime} is chiral, leptophobic and flavor non-universal.  
One can imagine that $Z'$ is associated with a new local gauge symmetry $U(1)'$. 
Then the original $Z'$ model has gauge anomalies and is mathematically inconsistent.  
Also one can not write Yukawa couplings for the up-type quarks if we have only the 
SM Higgs doublet which has the vanishing $U(1)'$ charge.  Then the top quark would
be massless, which is physically unrealistic and unacceptable. 
Therefore it would be highly nontrivial to construct a realistic gauge theory which 
satisfies the conditions in the original $Z'$ model.  
Let us recall that  the original $Z'$ model was excluded by the same sign top pair 
productions, because $Z'$ exchange can contribute to $ u u \rightarrow t t$.  
The  upper bounds on the same-sign top-pair production put
strong constraints on this model~\cite{saavedra}. 
However the model with extra $Z^{'}$ only is not either consistent or realistic 
because of the reasons described above.   The original $Z^{'}$ model should be 
extended with new Higgs doublets before one starts working on detailed phenomenology,
as described in the next section~\cite{u1models}. 

\subsection{$U(1)'$ models with flavored multi-Higgs doublets by Ko, Omura and Yu
~\cite{u1models}}

In this subsection we review the flavor-dependent chiral U(1)$^\prime$ model 
with flavored Higgs doublets that was proposed in Ref.~\cite{u1models}. 
Our model is an extension of the $Z^\prime$ model ~\cite{zprime} described in the 
previous section, curing various problems of Ref.~\cite{zprime}.  
The $Z^\prime$ boson must be associated with some gauge symmetry 
if we work in weakly interacting theories, and we consider an extra leptophobic 
U(1)$^\prime$ symmetry~\cite{u1models}.  
And in order to avoid too large FCNCs in the down quark sector, we assigned 
flavor-dependent  U(1)$^\prime$ charges $u_i$ $(i=u,c,t)$
only to the right-handed up-type quarks while the left-handed quarks and
right-handed down-type quarks are not charged under $U(1)'$.  

Then, the Lagrangian between $Z^\prime$ and the SM quarks 
in the interaction eigenstates is given by
\begin{equation}
\mathcal{L}_{Z^\prime q\bar{q}} =
g^\prime \sum_i u_i  Z^\prime_\mu \overline{U_R^i} \gamma^\mu U_R^i,
\end{equation}
where $U_R^i$ is a right-handed up-type quark field in the interaction eigenstates 
and $g^\prime$ is the couping of the U(1)$^\prime$. 

After the electroweak symmetry breaking, we can rotate the quark fields 
into the mass eigenstates by bi-unitary transformation.   
The interaction Lagrangian for the $Z^\prime$ boson in the mass eigenstate
is given by
\begin{equation}
\mathcal{L}_{Z^\prime q\bar{q}}   =    
g^\prime Z^\prime_\mu \left[  
(g_R^u)_{ut} \overline{u_R} \gamma^\mu t_R
+(g_R^u)_{ut} \overline{t_R} \gamma^\mu u_R
+ (g_R^u)_{uu} \overline{u_R} \gamma^\mu u_R
+(g_R^u)_{tt} \overline{t_R} \gamma^\mu t_R
\right].
\end{equation}
The $3\times 3$ mixing matrix $(g_R^u)_{ij} = (R_u)_{ik} u_k (R_u)^\dagger_{kj}$ 
is the product of the U(1)$^\prime$ charge matrix ${\rm diag} ( u_{k=1,2,3} )$ and
a unitary matrix $R_u$, where the matrix  $R_u$ relates the RH up-type quarks 
in the interaction eigenstates and in the mass eigenstates.   
The matrix $R_u$ participates in diagonalizing the up-type quark mass matrix. 
We note that the components of the mixing angles related 
to the charm quark have to be small in order to respect 
constraints from the $D^0$-$\overline{D^0}$ mixing.

If one assigns the U(1)$^\prime$ charge $(u_i)=(0,0,1)$
to the right-handed up-type quarks, one can find the relation 
$(g_R^u)_{ut}^2 = (g_R^u)_{uu} (g_R^u)_{tt}$ \footnote{We note 
that the relation is not valid for the other charge assignments. For general cases, 
we introduce a parameter $\xi$ with $(g_R^u)_{uu} (g_R^u)_{tt} = \xi (g_R^u)_{ut}^2 $
where $\xi$ is a free parameter. }. 
This relation indicates that if the $t$-channel diagram mediated by $Z^\prime$ 
contributes to the $u\bar{u}\to t\bar{t}$ process, the $s$-channel diagram 
mediated by $Z^\prime$ should be taken into account, too.

As we discussed in the previous section, it is mandatory to include
additional flavored Higgs doublets charged under U(1)$^\prime$ in order 
to write down proper Yukawa interactions for the SM quarks charged
under U(1)$^\prime$ at the renormalizable level \footnote{It is also true that 
one cannot write nonrenormalizable Yukawa interactions with  the SM 
Higgs doublet only. It is essential to include the Higgs doublets with nonzero 
$U(1)^{'}$ charges in order that one can write Yukawa couplings for the up-type
quarks in this model.}.
The number of additional Higgs doublets depends on the U(1)$^\prime$ charge 
assignment to the SM fermions,  especially the right-handed up-type quarks. 
In general, one must add three additional Higgs doublets with U(1)$^\prime$ 
charges $u_i$ (see Ref.~\cite{u1models} for more discussions).
For the charge assignment $(u_i)=(0,0,1)$ we have two Higgs doublets
including the SM-like Higgs doublet, while for $(u_i)=(-1,0,1)$ three Higgs
doublets are required. The additional U(1)$^\prime$ must be broken in the end, 
so that we add a U(1)$^\prime$-charged singlet Higgs field $\Phi$ to the SM. 
Both the U(1)$^\prime$-charged Higgs doublet and the singlet $\Phi$ can give
the masses for the $Z^\prime$ boson and extra fermions if it has
a nonzero vacuum expectation value (VEV). After breaking of the electroweak 
and U(1)$^\prime$ symmetries, one can write down the Yukawa interactions 
in the mass basis.
After all the Yukawa couplings would be proportional to the quark 
masses responsible for the interactions so that we could ignore 
the Yukawa couplings which are not related to the top quark. 

The number of relevant Higgs bosons participating in the top-quark pair 
production depends on the U(1)$^\prime$ charge assignment and mixing angles. 
The relevant Yukawa couplings for the top-quark pair production can be written as
\begin{equation}
V =  Y_{tu}^h \overline{u_L} t_R h+Y^H_{tu} \overline{u_L} t_R H 
+ i Y_{tu}^a \overline{u_L} t_R a + h.c.,
\end{equation}
where $h$ and $a$ are the lightest neutral scalar and pseudoscalar 
Higgs bosons, and $H$ is the heavier (second lightest) neutral Higgs boson.
We assume that the Yukawa couplings of the other Higgs bosons 
are suppressed by the mixing angles \footnote{
This assumption is not compulsory, since all the Higgs bosons might 
participate in the top-quark pair production in principle. We will keep only
a few lightest (pseudo) scalar bosons in order to simplify the numerical analysis. 
}.

Introducing $U(1)'$ flavored Higgs doublets is very important because they generate  
nonzero top mass.  They also play an important role in top FBA phenomenology.  
For example the Yukawa couplings of the neutral scalar bosons $h,H,a$ have flavor 
changing  couplings to the up-type quarks because of the flavor non-universal nature 
of $Z'$ interaction~\cite{u1models}:
\begin{eqnarray} 
Y^h_{tu} & = & \frac{ 2m_t (g^u_R)_{ut} }{v \sin( 2 \beta)} 
\sin (\alpha-\beta) \cos \alpha_{\Phi} \ ,  
\\
Y^H_{tu} & = & - \frac{ 2m_t (g^u_R)_{ut} }{v \sin( 2 \beta)} 
\cos (\alpha-\beta) \cos \alpha_{\Phi} \ ,
\\
Y^a_{tu} & = & \frac{ 2m_t (g^u_R)_{ut} }{v \sin( 2 \beta)} \ .
\end{eqnarray} 
These Yukawa couplings are not present in the Type-II 2HDM, for example.  
Our models  proposed in Ref.~\cite{u1models} are good examples of non-minimal 
flavor violating multi-Higgs doublet  models, where the non-minimal flavor violation 
originates from the flavor non-universal chiral couplings of the new gauge 
boson $Z'$.   In our model,  the top FBA and the same-sign top-pair productions are 
generated not only by the $t$-channel $Z'$ exchange, but also by the $t$-channel 
exchange of neutral Higgs scalars,  and the strong constraint on the original $Z'$ 
model  from the same-sign top-pair production can be relaxed by a significant 
amount when we include all the contributions in the model, as described 
in the following section. 

\subsection{Phenomenology}

In this subsection, we discuss phenomenology of our model described in the previous 
subsection. If new physics affects the top-quark pair production and could accommodate  
$A_\textrm{FB}^t$ at the Tevatron,  it must also be consistent with many other 
experimental measurements related with the top quark.
In our models, both the $Z^\prime$ and Higgs bosons $h$ and $a$  contribute
to the top-quark pair production through the $t$-channel exchange in 
the $u\bar{u} \to t \bar{t}$ process.  As we discussed in the previous section,
the $Z^\prime$ boson also  contributes to the top-quark pair production 
through the $s$-channel exchange, which was ignored in Ref.~\cite{zprime}.

As two extreme cases, one can consider the cases where only the $Z^\prime$ boson
or Higgs boson $h$ contributes to the top-quark pair production.
Then, our models become close to the simple $Z^\prime$ model of Ref.~\cite{zprime}
or the scalar-exchange model of Ref.~\cite{babu}. 
Unfortunately, these models cannot be compatible with the present upper bound 
on the same-sign top-quark pair production at the LHC in the parameter space 
which give rise to a moderate $A_\textrm{FB}^t$~\cite{u1models}. 
In our chiral U(1)$^\prime$ models, the constraint from the same-sign top-quark
pair production could be relaxed because of the destructive interference
between the contribution from the $Z^\prime$ and those from Higgs bosons $h$ 
and $a$.  In particular, the contribution of the pseudoscalar boson $a$ to the 
same-sign  top-quark pair production is opposite to the other contributions.

In the two Higgs doublet model with the $U(1)^\prime$ assignments  
to the right-handed up-type quarks, $(u_i)=(0,0,1)$,  
the $s$-channel contribution of the $Z^\prime$ exchange 
to the partonic process  $u\bar{u}\to t\bar{t}$  is as strong as 
an $t$-channel contribution because of the relation 
$(g_R^u)_{ut}^2 = (g_R^u)_{uu} (g_R^u)_{tt}$~\cite{u1models}.
In the multi-Higgs doublet models (mHDMs)  with other U(1)$^\prime$ charge 
assignments $(u_i)'$s to the right-handed up-type quarks, 
the $s$-channel contribution could be small.
In general, one can write $(g_R^u)_{uu} (g_R^u)_{tt} = \xi (g_R^u)_{ut}^2$,
where $\xi$ is a function of mixing angles and $0\leq |\xi| \leq O(1)$. 
In the case of $m_{Z^\prime} \geq 2 m_t$, a resonance around the $Z^\prime$ 
mass for nonzero $\xi$ would be observed in the $t\bar{t}$ invariant mass 
distribution.  However, such a resonance has not been observed so far in the experiments.  This would restrict the $Z^\prime$ mass to be 
much smaller than $2 m_t$ for nonzero $\xi$.
For the numeric values for $\sigma^{t\bar{t}}$, $\sigma^{tt}$, $A_C^y$ and
other related data and parameters, I refer to the original paper~\cite{Ko:2012ud}.

{\it (i) $m_{Z'} = 145$ GeV and $\xi=1$}: 
In this case, the $Z^\prime$ boson can contribute to the top-quark pair
production through its $s$-channel and $t$-channel exchanges
in the $u\bar{u}\to t\bar{t}$ process. While the Higgs bosons contribute
to the top-quark pair production
only in the $t$ channel because the diagonal elements of their Yukawa couplings 
to light quarks   are negligible.
We scan the following parameter regions:
$180~\textrm{GeV} \le m_{H,a} \le 1~\textrm{TeV}$,
$0.005 \le \alpha_x \le 0.012$,
$0.5 \le Y_{tu}^{H,a} \le 1.5$, and
$(g_R^u)_{tu}^2=(g_R^u)_{uu} (g_R^u)_{tt}$, 
where $\alpha_x \equiv (g_R^u)_{tu}^2 g'^2/(4 \pi)$ is defined and $Y_{tu}^{H,a}$ are flavor-off-diagonal Yukawa couplings. 

In Fig.~1, we show the scattered plot for $A_\textrm{FB}^t$ at the Tevatron and 
the same-sign top-pair production cross section and $A_C^y$ 
at the LHC. 
The green and yellow regions are consistent with $A_C^y$ at ATLAS and CMS
in the $1\sigma$ level, respectively. The blue and skyblue regions are
consistent with $A_{FB}^t$ in the lepton+jets channel at CDF in the
$1\sigma$ and $2\sigma$ levels, respectively.
The red points are in agreement with the cross section 
for the top-quark pair production
at the Tevatron in the $1\sigma$ level and the blue points are consistent with
both the cross section for the top-quark pair production at the Tevatron
in the $1\sigma$ level and the upper bound on the same-sign top-quark pair
production at ATLAS. We find that a lot of parameter points can explain all
the experimental data.
We emphasize that the simple $Z^\prime$ model is excluded by the same-sign 
top-quark pair production, but in the chiral U(1)$^\prime$ model, this strong bound 
could be evaded  due to the destructive interference between the $Z^\prime$ boson 
and Higgs bosons.  Also the $m_{t\bar{t}}$ distribution becomes closer to the SM 
case in the presence of $h$ and $a$ contributions (see Fig.~2).  One can realize that it is important to include the Higgs contributions 
 as well as the $Z'$ contributions. All the physical observables are affected by the Higgs contributions.

\begin{figure}[!t]
\epsfig{file=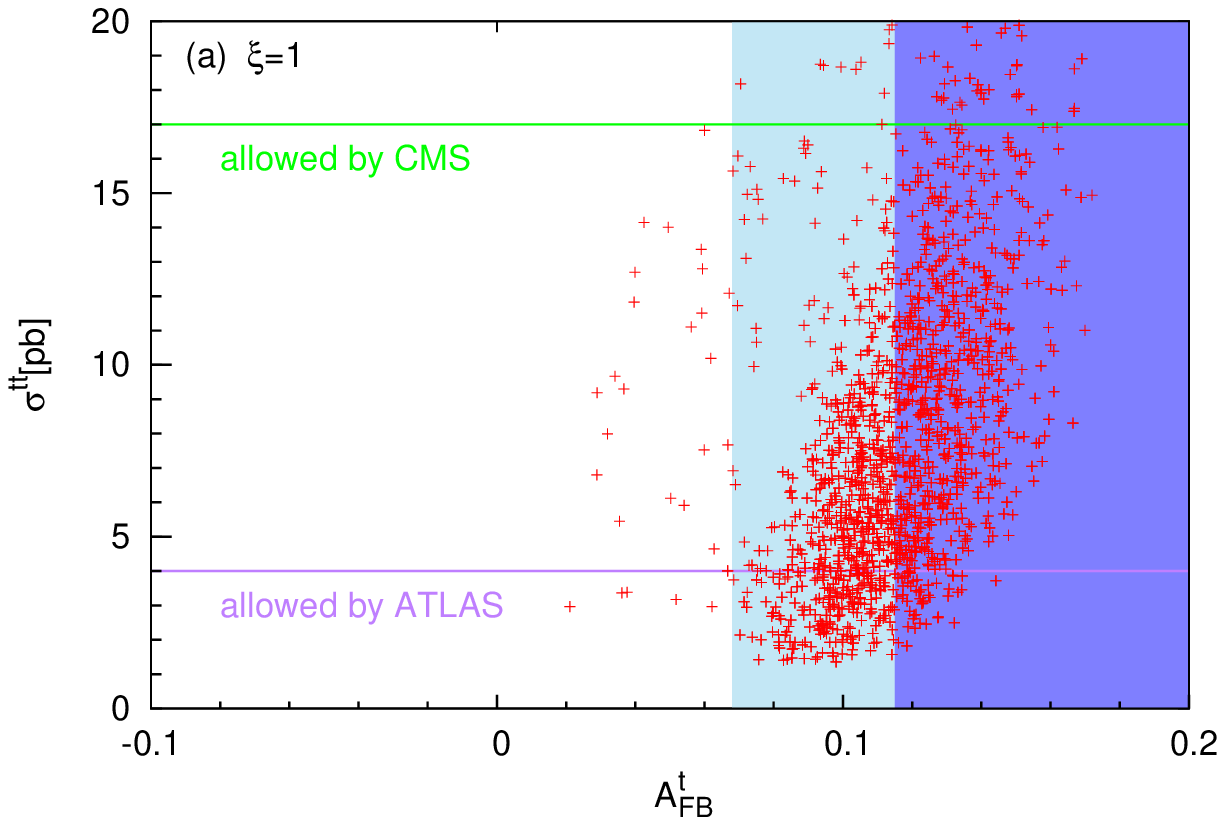,width=0.45\textwidth}
\epsfig{file=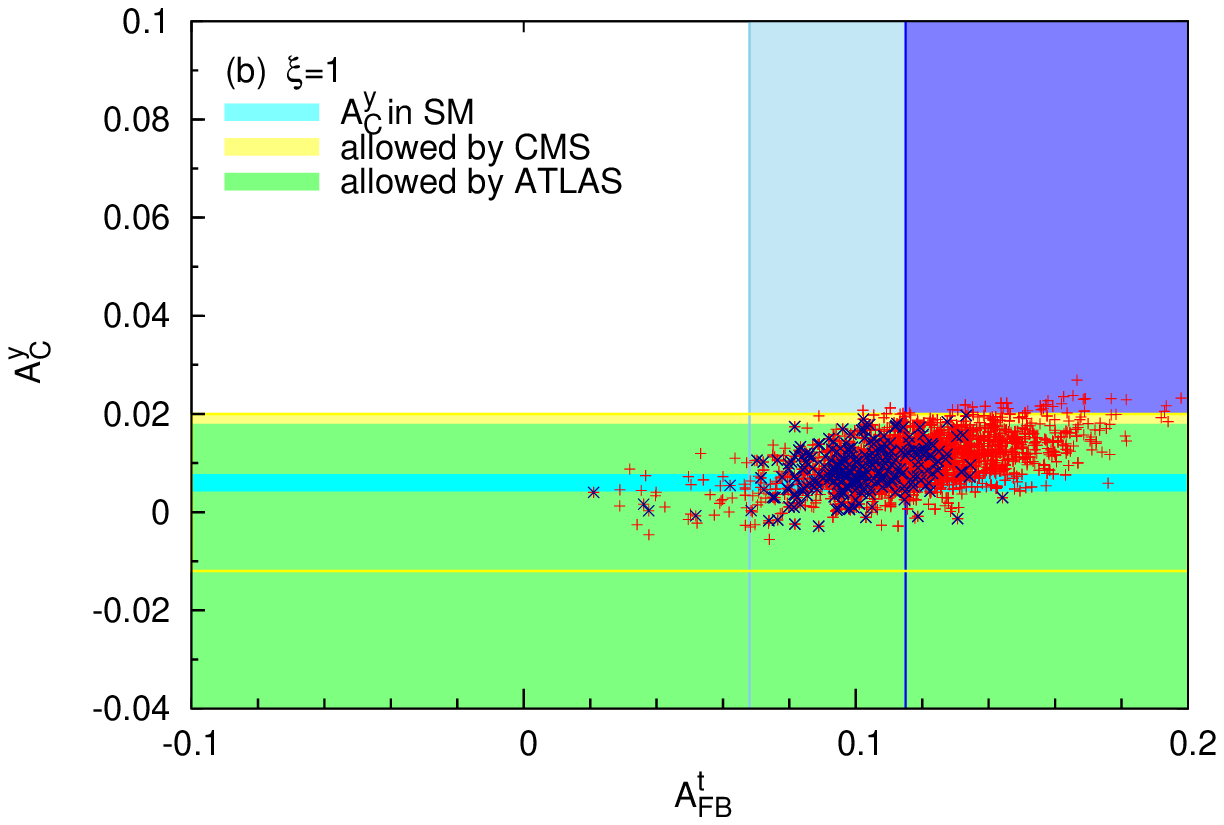,width=0.45\textwidth}
\caption{\label{fig:lightzp}%
The scattered plots for (a) $A_\textrm{FB}^t$ at the Tevatron and
$\sigma^{tt}$ at the LHC in unit of pb, and (b) $A_\textrm{FB}^t$ at the 
Tevatron and $A_C^y$ at the LHC for $m_{Z^\prime}=145$ GeV and $\xi=1$.
In (b), the blue points satisfy the upper bound on the same sign top pair 
production from ATLAS: $\sigma^{tt} < 4$ pb.
}
\end{figure}

\begin{figure}[!t]
\epsfig{file=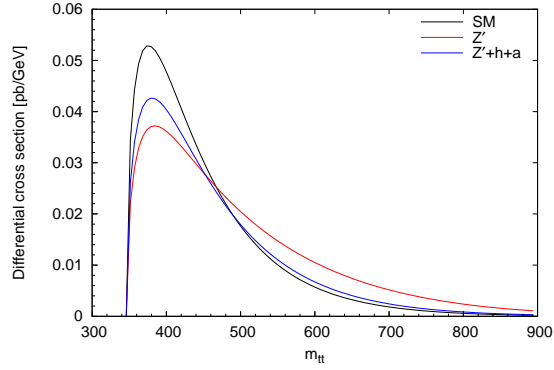,width=0.45\textwidth}
\caption{\label{fig:mtt}%
The invariant mass distribution of the top-quark pair at the Tevatron
in the SM, $Z^\prime$ model, and chiral U(1)$^\prime$ model.
}
\end{figure}

{\it (ii) $m_h = 125$ GeV and $\xi=0$}: 
In this case, we discuss the scenario that
 a light Higgs boson $h$ with $m_h=125$ GeV, motivated by the recent observation 
 of an SM-Higgs like scalar boson 
at the LHC~\cite{higgs}, also has a nonzero $Y^h_{tu}$. 
In this case, the $Z^\prime$ boson and Higgs bosons
$h$, $H$, and $a$ contribute to the top-quark pair production.
In order to suppress the exotic decay of the top quark into $h$ and $u$, 
we set the Yukawa
coupling of $h$ to be $Y_{tu}^h \le 0.5$ and masses of $Z^\prime$, $H$,
and $a$ are larger than the top-quark mass or approximately equal to the
top-quark mass. We scan the following parameter regions:
$160~\textrm{GeV} \le m_{Z^\prime} \le 300~\textrm{GeV}$,
$180~\textrm{GeV} \le m_{H,a} \le 1~\textrm{TeV}$,
$0 \le \alpha_x \le 0.025$,
$0 \le Y_{tu}^{H,a} \le 1.5$, $0 \le Y_{tu}^{h} \le 0.5$ and $\xi=0$. 
The mass region of the $Z^\prime$ boson
is taken to avoid the constraint from the $t\bar{t}$ invariant mass
distribution at the LHC. If $(g_R^u)_{uu}\simeq0$ 
and the $s$-channel contribution of the $Z^\prime$ could be ignored,
the mass region of the $Z^\prime$ boson could be enlarged.
In Fig.~3, we show the scattered plot for $A_\textrm{FB}^t$ 
at the Tevatron and $A_C^y$ at the LHC for $m_h=125$ GeV.
All the legends on the figure are the same as those in Fig.~1.
We find that there exist parameter regions which agree with all the experimental
constraints. 
We emphasize that in some parameter spaces $\sigma^{tt}$ is less than 1 pb.

\begin{figure}[!th]
\epsfig{file=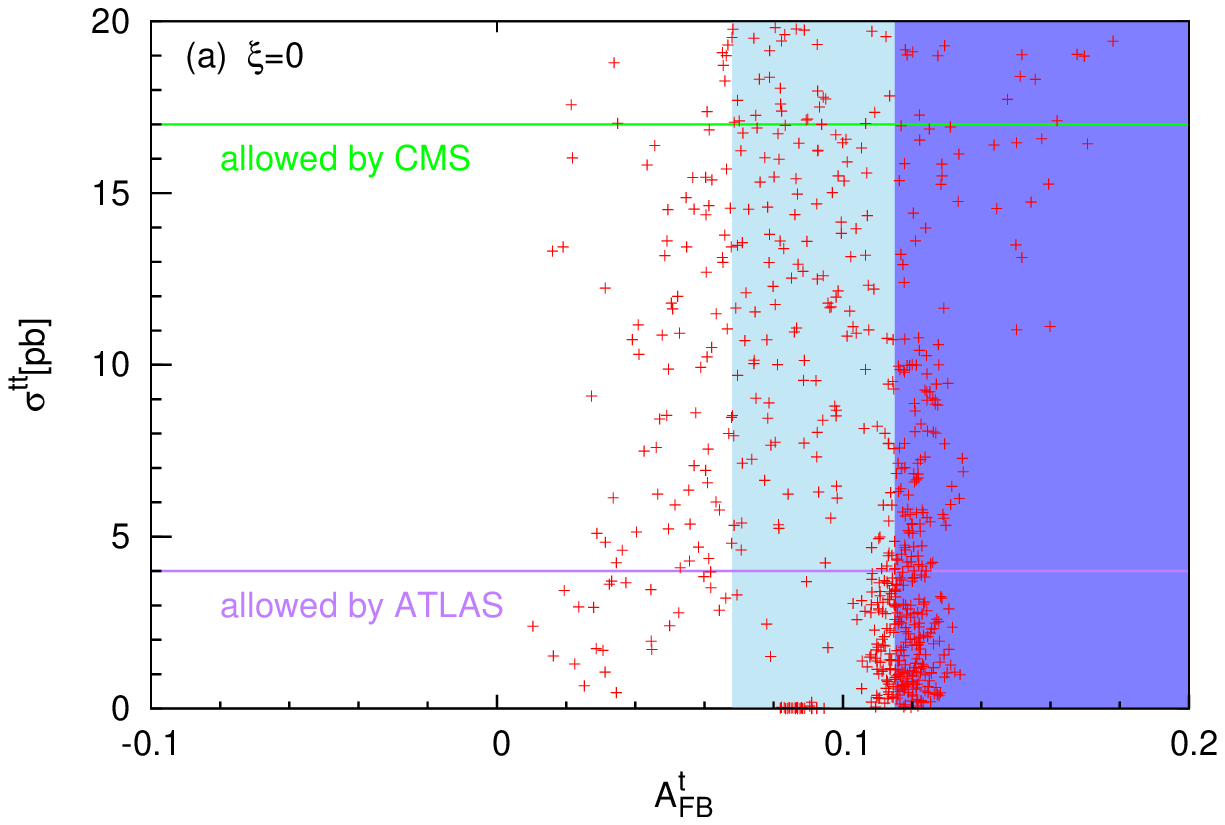,width=0.45\textwidth}
\epsfig{file=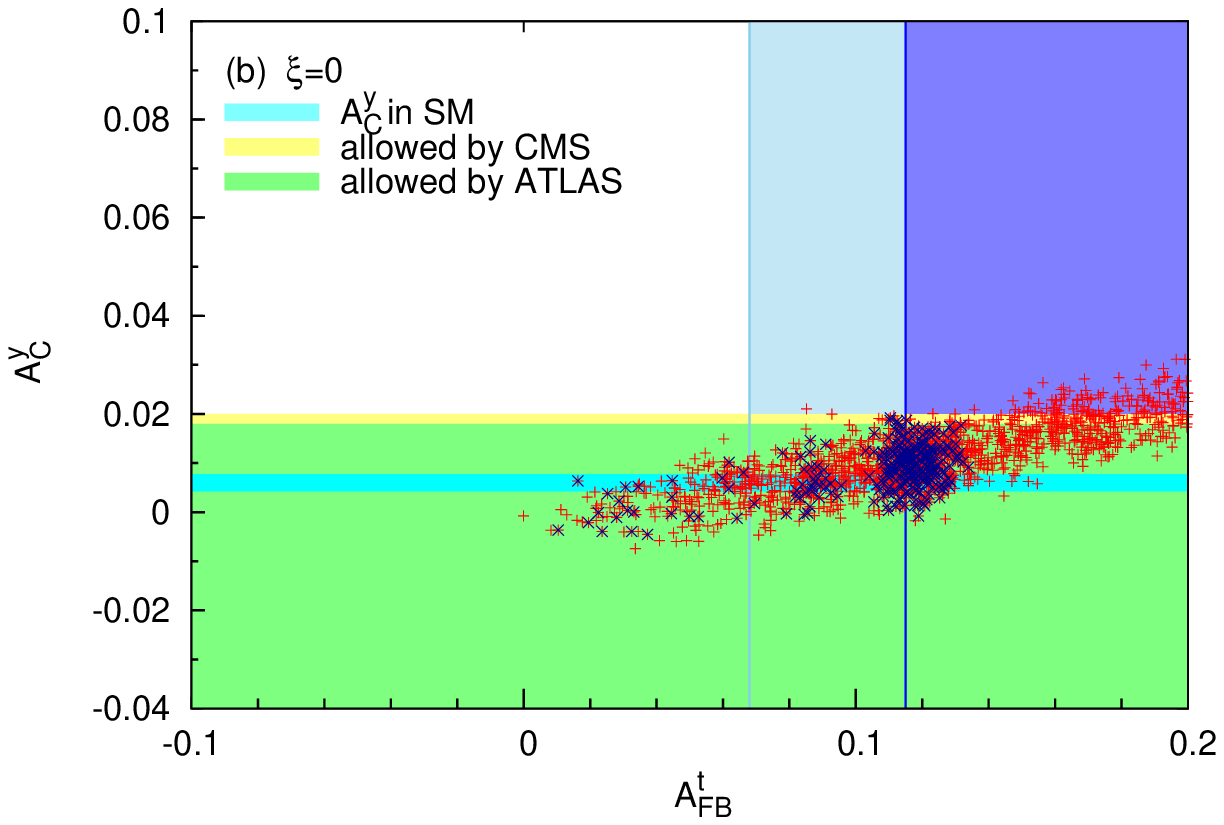,width=0.45\textwidth}
\caption{\label{fig:higgs2-2}%
The scattered plots for (a) $A_\textrm{FB}^t$ at the Tevatron and
$\sigma^{tt}$ at the LHC in unit of pb, and (b) $A_\textrm{FB}^t$ at the 
Tevatron and $A_C^y$ at the LHC for $m_h=125$ GeV and $\xi=0$,
where the contribution of the second lightest Higgs boson $H$ is included.
}
\end{figure}

\subsection{Summary}
The top forward-backward asymmetry at the Tevatron is the only quantity which has 
deviation from the SM prediction in the top quark sector up to now.
A lot of new physics models have been introduced to account for this deviation.  
Or it has been analyzed in a model-independent way~\cite{modelindep,grojean}, 
and some models have already been disfavored by experiments at the LHC.
In this section, we investigated the chiral U(1)$^\prime$ model with flavored
Higgs doublets and flavor-dependent couplings. Among possible scenarios,
we focused on two scenarios,  
both of which  can accommodate with the constraints from the same-sign top-quark 
pair production and the charge asymmetry at the LHC as well as the top 
forward-backward asymmetry  at the Tevatron.

The chiral U(1)$^\prime$ model has a lot of new particles except for
the $Z^\prime$ boson and neutral Higgs bosons. The search for exotic particles
may constrain our model severely. For example, our model is strongly
constrained by search for the charged Higgs boson in the
$b\to s\gamma$, $B\to \tau \nu$, and $B\to D^{(\ast)}\tau \nu$ decays \cite{bdecay}.
In order to escape from such constraints, we must assume a quite heavy charged
Higgs boson or it is necessary to study our model more carefully
by including all the interactions which have been neglected in this work.
More detailed analysis on this issue can be found in Ref.~\cite{bdecay}.

\section{A new resolution of Higgs-mediated FCNC in 2HDMs 
with local $U(1)_H$ Higgs flavor symmetry}

\subsection{Preamble}
Adding one more Higgs doublet to the SM is one of the simplest 
extensions of the SM, leading to the so-called two Higgs doublet model (2HDM). 
The 2 HDM's have been studied in various contexts (see, for example,
Ref.~\cite{Branco:2011iw} for a recent review).
Generic 2HDM's suffer from excessive flavor changing neutral current 
(FCNC) mediated by neutral Higgs boson exchanges. 
This is due to the fact that the individual Yukawa couplings would not be diagonalized
simultaneously when the fermion mass matrices are diagonalized by unitary matrices
acting on the left-handed and the right-handed quarks and leptons.  

One way to avoid this problem is to impose an ad hoc $Z_2$ discrete 
symmetry as suggested by Glashow and Weinberg long time ago 
\cite{Glashow:1976nt}, 
which is often called Natural Flavor Conservation (NFC) : 
\[
Z_2 : ( H_1 , H_2 ) \rightarrow (+H_1 , - H_2 ) .
\]
The Yukawa sectors can be controlled by assigning suitable $Z_2$
parities to the SM fermions, and the models are often categorized 
into four types (see Table 1) ~\cite{Barger:1989fj,Aoki:2009ha}:
\begin{table}[th]
\begin{tabular}{cccccccc}
\hline 
Type & $H_1$ & $H_2$ & $U_R$ & $D_R$ & $E_R$ & $N_R$ & $Q_L, L$
 \\
 \hline
  I  & $+$   & $-$   & $+$   & $+$   & $+$   & $+$   & $+$ 
 \\
 II  & $+$   & $-$   & $+$   & $-$   & $-$   & $+$   & $+$
 \\
 X & $+$   & $-$   & $+$   & $+$   & $-$   & $-$   & $+$
 \\
 Y  & $+$   & $-$   & $+$   & $-$   & $+$   & $-$   & $+$    
 \\ \hline
\end{tabular}
\caption{}
\end{table}
However it is well known that discrete symmetry could generate a 
domain wall problem when it is spontaneously broken, which is indeed
the case in the 2HDM. Therefore the $Z_2$ symmetry is assumed to be 
broken softly by a dim-2 operator, $H_1^\dagger H_2$ term. 
Also the origin of such a discrete symmetry is not clear at all. 

For long time this Higgs-mediated FCNC problem was solved or evaded by assuming
the so-called Natural Flavor Conservation (NFC) criterion proposed by Glashow and 
Weinberg~\cite{Glashow:1976nt}.  
In practice, this criterion  amounts to assume that the fermions  of the same electric 
charges get their masses from only one type of Higgs doublet. 
This criterion is easily realized in the two-Higgs doublet model (2HDM) by imposing 
softly broken $Z_2$ symmetry under which $H_1$ and $H_2$ and the SM chiral fermions
are charged differently so that the NFC criterion by Glashow-Weinberg is realized. 
However the origin of the discrete $Z_2$ symmetry and its soft breaking is not clear 
at all.  

In Ref.~\cite{koy}, we proposed a new resolution of the Higgs mediated FCNC
problem in 2HDMs, by implementing the discrete $Z_2$ symmetry to local $U(1)_H$ 
Higgs flavor symmetry.  In the $U(1)_H$ extensions of the usual 2HDMs,  it is important
to impose the anomaly cancellation in addition to the phenomenologically viable 
Yukawa interactions, which controls the $U(1)_H$ quantum numbers of the SM fermions
as  well as new chiral fermions introduced for anomaly cancellation.  Therefore the 
$U(1)_H$ extension of 2HDMs are not really the same as the ordinary 2HDM, even in 
the limit of infinitely heavy $U(1)_H$ gauge boson.  Even if we  integrating out the $Z_H$ gauge boson assuming they are very heavy,  there could be remaining new chiral fermions 
which were necessary for the anomaly cancellation.  This simple resolution of the 
Higgs mediated FCNC problem was not proposed before, and it is worthwhile to 
study its phenomenology at colliders in more detail~\cite{typeI}, 
at low energy flavor physics as well as in the context of 
electroweak phase transition and baryogenesis. 

\subsection{Higgs sector}

Let us assume that $H_1$ and $H_2$ carry different $U(1)_H$ charges, 
$h_1$ and $h_2$ (with $h_1 \neq h_2$ in order to distinguish two of them), 
with $g_H$ being the $U(1)_H$ coupling.  The kinetic terms for the $H_1$ and 
$H_2$ will involve the $U(1)_H$ couplings:
\begin{equation}
D_\mu H_i = D_\mu^{\rm SM} H_i - i g_H h_i Z_{H \mu} H_i
\end{equation}
with $i=1,2$. 
Then the mass matrix for $Z$ and $Z_H$ from the kinetic terms of
$H_1$ and $H_2$ is given by 
\begin{equation}
M^2 = \left( \begin{array}{cc}
                    g_Z^2 v^2   &   -g_Z g_H ( h_1 v_1^2 + h_2 v_2^2 )  \\
                    -g_Z g_H ( h_1 v_1^2 + h_2 v_2^2 )  & 
                    g_H^2 ( h_1^2 v_1^2 + h_2^2 v_2^2 ) 
                    \end{array}   \right) \ ,
\end{equation}
where $v^2 = v_1^2 + v_2^2$. 
Note that the determinant of $M^2$ is not zero, as long as 
$h_1 \neq h_2$.
If we add an additional $U(1)_H$ charged singlet scalar $\Phi$ 
(its $U(1)_H$ charge is defined as $\phi$) with nonzero VEV 
$v_\phi$,  the $(22)$ component of the (mass)$^2$ matrix 
would have an additional piece  $g_H^2 h^2_{\Phi} v_\Phi^2$ from the 
kinetic term of $\Phi$.  The mass mixing must be small to avoid too large 
deviation of $\rho$ parameter  from the SM prediction. 
The tree-level deviation within $1\sigma$ 
restricts the mass and coupling of $Z_H$:
\begin{equation}
 \{ h_1 (\cos \beta )^2+ h_2 (\sin \beta )^2 \}^2
 \frac{ g^2_H}{ g^2_Z}\frac{  m_{\Hat{Z}}^2 }{ m_{\hat{Z}_H}^2 - m_{\Hat{Z}}^2}
 \lesssim O(10^{-3}),
\end{equation}
where $ m_{\Hat{Z}}^2 = g_Z^2 v^2$ and 
$ m_{\hat{Z}_H}^2= g_H^2  v^2 \{ h^2_1 (\cos \beta )^2+h^2_2 (\sin \beta )^2 \}
+g_H^2  h^2_{\Phi} v_\Phi^2  $.

The potential of our 2HDM is given by 
\begin{equation}
V (H_1,H_2) =  m_1^2 H_1^\dagger H_1 + m_2^2 H_2^\dagger H_2 
+ \frac{\lambda_1}{2} ( H_1^\dagger H_1 )^2 
+ \frac{\lambda_2}{2} ( H_2^\dagger H_2 )^2 
+ \lambda_3 H_1^\dagger H_1 H_2^\dagger H_2 
+ \lambda_4 H_1^\dagger H_2 H_2^\dagger H_1 .
\end{equation}
In terms of the standard notation for the 2HDM potential, our 
model corresponds to a special case $m_3^2 = \lambda_5 = 0$. 
Note that $H_1^\dagger H_2$ or its square are forbidden by $U(1)_H$ 
symmetry, since we have imposed $h_1 \neq h_2$.
If the model were not gauged with the extra $U(1)_H$, one would
encounter the usual problem of a massless pseudoscalar $A$.
In our case, this massless mode is eaten by the $U(1)_H$ gauge boson,
and there is no usual problem with a massless Goldstone boson.
Instead the scalar boson spectrum is different from the usual 
2HDM, since there would no pseudoscalar $A$ in our models.

In case we include a singlet scalar $\Phi$, let us define $\phi = h_1 - h_2$, 
so that $H_1^\dagger H_2 \Phi$ is gauge invariant.  
Then there would be additional terms in the scalar potential:
\begin{equation}
\Delta V = m_\Phi^2 \Phi^\dagger \Phi
+ \frac{\lambda_\Phi}{2} (\Phi^\dagger \Phi )^2
+ ( \mu H_1^\dagger H_2 \Phi + H.c. ) 
+\mu_1 H_1^\dagger H_1 \Phi^\dagger \Phi
+ \mu_2 H_2^\dagger H_2 \Phi^\dagger \Phi , 
\end{equation}
depending on $h_1$ and $h_2$ and $\Phi$. 
After $\Phi$ develops a VEV, $\mu$ terms look like the 
$m_3^2$ term in the conventional notation. And the effective 
$\lambda_5$ term is generated by the $\Phi$ mediation: 
$\lambda_5 \sim \mu^2 / m_\Phi^2$ well below $m_\Phi$ scale.
In any case there is no dangerous Peccei-Quinn symmetry leading 
to a massless $Z^0$ unlike the usual 2HDM, and no need for
soft breaking of $Z_2$ symmetry, because of extra $U(1)_H$ 
gauge symmetry.

Production and decay modes of the new $Z_H$ gauge boson will depend on 
the $U(1)_H$ charges of the SM fermions, which will differ case by case.
In the following, we implement each 2HDM's  
with NFC (Type-I,II,X,Y) to local 
$U(1)_H$ gauge theories by assigning suitable $U(1)_H$  charges to two Higgs
doublets $H_1$ and $H_2$ and the SM fermions, and by adding new chiral 
fermions for anomaly cancellation.

\subsection{Type-I 2HDM\label{subsec:type1}}

Let us first start with the simplest case, the Type-I 2HDM, where 
the SM fermions can get masses only from $H_1$ VEV. 
This is possible, if (with $h_1 \neq h_2$)
\begin{equation}
u - q  - h_1 = d - q + h_1 = e - l + h_1 = n - l - h_1 = 0 .
\end{equation}
There are many ways to assign $U(1)_H$ charges to the SM fermions
to achieve this scenario.  The phenomenology will depend crucially 
on the $U(1)_H$ charge  assignments of the SM fermions. 
In general, the models will be anomalous, even if $U(1)_H$ charge 
assignments are nonchiral, so that one has to achieve anomaly 
cancellation by adding new chiral fermions to the particle spectrum. 

For the Type-I case, one can achieve an anomaly-free $U(1)_H$ 
assignment even without additional chiral fermions: 
\begin{table}[th]
\begin{tabular}{cccccccc}
\hline 
Type & $U_R$ & $D_R$ & $Q_L$ & $L$ & $E_R$ & $N_R$ & $H_1$ 
\\
\hline
$U(1)_H$ charge & $u$   & $d$   & $\frac{(u+d)}{2}$ & $\frac{-3 (u+d)}{2}$ & 
$-(2 u + d)$ & $-(u+2d)$ & $ \frac{(u-d)}{2}$
\\
\hline
$h_2 \neq 0$ & $0$   & $0$    & $0$   & $0$    & $0$  & $0$  & $0$  
\\
$U(1)_{B-L}$ & $1/3$ & $1/3$  & $1/3$ & $-1$   & $-1$ & $-1$ & $0$ 
\\
$U(1)_R$  & $1$   & $-1$   & $0$   & $0$    & $-1$ & $1$  & $1$  
\\
$U(1)_Y$ & $2/3$ & $-1/3$ & $1/6$ & $-1/2$ & $-1$ & $0$  & $ 1/2$
 \\ \hline
\end{tabular}
\caption{Anomaly free $U(1)_H$ assignments in the Type-I 2HDM}
\end{table}
There is one free parameter by which the charge assignments determines
the theory, modulo the overall coupling constant $g_H$. It is amusing 
to observe that there appear an infinite number of new models which is 
a generalization of the Type-I model into Higgs flavor $U(1)_H$ models 
without extending the fermion contents at all. 

There are four simple and interesting anomaly-free charge assignments 
without new chiral fermions, however: 
\begin{itemize}
\item $(u,d) = (0,0)$: 
In this case, all the SM fermions are $U(1)_H$ singlets.
Then $Z_H$ is fermiophobic and Higgsphilic. It would not be easy
to find it at colliders because of this nature of $Z_H$, and $h_2 \neq 0$. 
In this case, $H^\pm W^\mp Z_H$ couplings from the Higgs kinetic 
terms would be the main source of production and discovery for $Z_H$. 
The phenomenology of $Z_H$ will be similar to the leptophobic 
$Z'$ studied in Ref.~\cite{Georgi:1996ei}.
\item $(u,d) = (\frac{1}{3}, \frac{1}{3})$ :
In this case, we have $U(1)_H = U(1)_{B-L}$, and 
$Z_H$ is the $(B-L)$ gauge boson, which gets  
mass from the doublet $H_2$ (and also by a singlet $\Phi$, if 
we include it).
Our case is very different from the usual $(B-L)$ model where 
$U(1)_{B-L}$ is broken only by the SM singlet scalar $\Phi$.
Therefore the phenomenology would be very different. 
However the Yukawa sector is controlled by $U(1)_H$ and a new 
Higgs doublet $H_2$ with nonzero $U(1)_H$ charge $h_2$.  
\item $(u,d) = (1, -1)$ :
In this case, we have $U(1)_H = U(1)_R$. 
The $Z_H$ couples only to the RH fermions, not to the LH fermions.
In this case, the would-be SM Higgs doublet $H_1$ also carries 
nonzero $U(1)_H$ charge, and Higgs phenomenology of this
type of models will be very different from the SM Higgs boson.
\item $(u,d) =  (\frac{2}{3}, -\frac{1}{3})$ : 
This case corresponds to $U(1)_H = U(1)_Y$, but it is different 
from the SM, since $h_1 = 1/2 \neq 0$, unlike the SM case 
where $h_1 = 0$.
Higgs phenomenology of this type of models will be very 
different from the SM Higgs boson.
\end{itemize}
Other interesting possibilities with vectorlike $U(1)_H$ are
to identify $U(1)_H = U(1)_B$ or $U(1)_L$. In these cases, 
however, the model becomes anomalous, and we have add 
additional chiral fermions for anomaly cancellation. 
Again, it is interesting to break $U(1)_B$ or $U(1)_L$ by
an $SU(2)_L$ doublet $H_2$ (and possibly by $\Phi$ too).

\subsection{Type-II 2HDM \label{subsec:type2}}

In this subsection, we will implement the Type-II model to a 
$U(1)_H$ gauge theory.
In the Type-II 2HDM, $H_1$ couples to the up-type fermions, 
while $H_2$ couples to the down-type fermions: 
\begin{equation}\label{eq;type2}
V_y = y_{ij}^U \overline{Q_{Li}} \widetilde{H_1} U_{Rj}
      + y_{ij}^D \overline{Q_{Li}} H_2 D_{Rj}
    + y_{ij}^E \overline{L_i} H_2 E_{Rj}
      + y_{ij}^N \overline{L_i} \widetilde{H_1} N_{Rj}.
\end{equation}

There could be a number of ways to achieve anomaly cancellation.
In this talk, I discuss only one possibility, relegating to our original paper for other 
interesting cases~\cite{koy}.  If we assign $(q,u,d)=(-1/3,2/3,-1/3)$ and $(l,e,n)=(0,0,1)$,
the $U(1)_H$-extended Type-II 2HDM corresponds to 
the leptophobic $Z^{'}$ model in the context of $E_6$ \cite{Rosner:1996eb},
with the following identification of $U(1)_H$ charge in terms of $U(1)$ 
generators of $E_6$ model: 
\[
Q_H = I_{3R} - Y_L + \frac{1}{2} Y_R .
\] 
The extra chiral fields introduced in Ref. ~\cite{Rosner:1996eb} 
cancel the anomaly:
\begin{table}
\begin{tabular}{ccccccc}\hline
         & $SU(3)$ & $SU(2)$ & $U(1)_Y$ & $U(1)_H$ \\ \hline 
$q_{Li}$ & $3$     & $1$     & $-1/3$   & $2/3$    \\ 
$q_{Ri}$ & $3$     & $1$     & $-1/3$   & $-1/3$   \\ 
$l_{Li}$ & $1$     & $2$     & $-1/2$   & $0$      \\ 
$l_{Ri}$ & $1$     & $2$     & $-1/2$   & $-1$     \\ 
$n_{Li}$ & $1$     & $1$     & $0$      & $-1$     \\ \hline      
\end{tabular}
\caption{Leptophobic $E_6$-type $U(1)_H$ charge assignments in Type-II 2HDM}
\end{table}
and the qualitative predictions made in Ref.~\cite{Rosner:1996eb} will 
apply in our case too without any modification.
Each generation has one extra vectorlike neutrinos (from $l_L$ and $l_R$ in Table~3) 
and one LH singlet neutrino ($n_L$), some of which has leptophobic gauge interaction.
Therefore baryonic neutrinos are realized in this model. 

The Yukawa couplings for SM fermions are given by the Eq.~(\ref{eq;type2}),
and the mass and mixing terms of the extra fermions will be generalized to 
\begin{eqnarray}
  V_m &=& y_{ij}^n \overline{n_{Li}} H_2 l_{Rj}
        + y_{ij}^q \overline{q_{Li}} q_{Rj} \Phi
        + y_{ij}^l \overline{l_{Li}} l_{Rj} \Phi
      + Y_{ij}^q \overline{Q_{Li}} H_2 q_{Rj} 
        + Y_{ij}^E \overline{l_{Li}} H_2 E_{Rj}
        \nonumber  \\
       & + & Y_{ij}^N \overline{l_{Li}} \widetilde{H_1} N_{Rj} 
      + Y_{ij}^D \overline{q_{Li}} D_{Rj} \Phi
        + Y_{ij}^l \overline{L_i} l_{Rj} \Phi + H.c.. 
\end{eqnarray}
Under this charge assignment, corresponding to $E_6$, the mixing terms 
between the SM fermions and the extra fermions are allowed at tree level, 
so that their Yukawa coupling must be tuned to avoid
the strong constraints from FCNC processes.  
 
One can repeat the same procedures for Type-X and Type-Y, and we refer to 
the original paper~\cite{koy} for more detailed discussion on this extension, and I am 
going to discuss in brief the neutrino physics in the $U(1)_H$ extension of Type-II 2HDM,
which is nothing but the leptophobic $E_6$ model discussed by J.L. Rosner sometime ago.

\subsection{Neutrino Physics in Type-II with leptophobic $E_6$ fermion contents}
After the spontaneous breaking of $SU(2)_L \times U(1)_Y \times U(1)_H \rightarrow 
U(1)_{\rm em}$ by nonzero VEV's 
\[
\langle H_1 \rangle = ( 0 , v_1 / \sqrt{2} )  , \ \ \
\langle H_2 \rangle = ( 0 , v_2 / \sqrt{2} )  , \ \ \
 \langle \Phi \rangle = v_\Phi / \sqrt{2} , 
\]
one can contract the mass matrices for the charged fermions and the neutrinos. 
In the interaction eigenstate basis, the neutrino mass matrix  is given by 
\[
{\cal M}_{\rm neutrino} = \frac{1}{\sqrt{2}} ~\left( 
\begin{array}{ccccc}
0 & 0 & Y^l v_\phi & 0 & y^N v_1 \\
0 & 0 & y^l v_\phi & 0 & Y^N v_1 \\
Y^l v_\phi & y^l v_\phi & 0 & y^n v_2 & 0 \\
0 & 0 & y^n v_2 & 0 & 0 \\
y^N v_1 & Y^N v_1 & 0 & 0 & 0 
\end{array}
\right)
\]
in the basis $ ( \nu_L  ,  \nu^{'}_L ,  N_R^{~c} , n_L ,  n^{~c}_L  )$ for each generation. 
There are a number of new sterile neutrinos in this model, which would result in very 
rich phenomenology both in particle physics and in cosmology. The detailed analysis 
of neutrino sector in the Type-II 2HDM with local $U(1)_H$ gauge symmetry with 
leptophobic $E_6$ matter  contents will be presented elsewhere~\cite{koy_neutrino}.

\subsection{Conclusion}

Let me summarize my talk. 
In this talk, I discussed 2 different types of multi-Higgs doublet models 
with local $U(1)_H$ Higgs flavor symmetry under which the SM fields could be 
charged too.   One model is for the top FBA with flavor dependent $U(1)_H$ couplings 
and the other being generalized 2HDMs with flavor universal $U(1)_H$ couplings. 
The main motivation for extending the Higgs sector with new Higgs doublets with 
nonzero $U(1)_H$ charges was to write realistic Yukawa interactions for the SM fermions.  

The idea of gauging Higgs flavor with $U(1)_H$ gauge symmetry can be applied
to the ordinary 2HDMs by implementing the softly broken discrete $Z_2$ symmetry 
to spontaneously broken local $U(1)_H$ symmetry. 
By assigning different $U(1)_H$ charges to two Higgs doublets  
and adjusting the $U(1)_H$ charges  of SM fermions properly, one can easily 
realize the ``Natural Flavor Conservation" suggested by Glashow and 
Weinberg \cite{Glashow:1976nt}.   There are infinitely many ways to 
assign $U(1)_H$ charges compatible with NFC, unlike the common 
practice based on discrete $Z_2$ symmetries.  Our proposal for Type-II 
2HDM has vastly different consequences from the MSSM 2HDM. In the
MSSM, the supersymmetric parts of the Higgs potential is Type-II, 
but eventually becomes Type-III when the loop corrections involving 
trilinear couplings are included.  And Higgs-mediated flavor 
violation can be enhanced by significant amount, 
especially for the large $\tan\beta$ region. On the other hand,
our models for Type-II are based on $U(1)_H$ gauge symmetry 
which is spontaneously broken.  The Higgs mediated FCNC is not
enhanced much even if we include the loop effects, unlike the MSSM.
We believe our proposal newly opens a wide window for the 2 HDM's.  

The basic ideas presented in this letter could be readily applied to other cases,
for example, to multi-Higgs doublet models in order to control the flavor 
problem by new gauge symmetries associated with Higgs fields.



\begin{theacknowledgments}
This work is supported in part by NRF Research Grant 2012R1A2A1A01006053 
(PK and CY).   PK would like to thank CETUP* (Center for Theoretical Underground 
Physics and Related Areas), supported by the US Department of Energy under Grant 
No. DE-SC0010137 and by the US National Science Foundation under Grant No. 
PHY-1342611, for its hospitality and partial support during the 2013 Summer Program.
\end{theacknowledgments}



\bibliographystyle{aipproc}   

\bibliography{sample}

\IfFileExists{\jobname.bbl}{}
 {\typeout{}
  \typeout{******************************************}
  \typeout{** Please run "bibtex \jobname" to optain}
  \typeout{** the bibliography and then re-run LaTeX}
  \typeout{** twice to fix the references!}
  \typeout{******************************************}
  \typeout{}
 }



\end{document}